\documentclass[aps,prl,amssymb,twocolumn]{revtex4}
\usepackage{amsfonts,amsmath,amsbsy,color}
\usepackage[]{graphics,graphicx}
\usepackage[version=3]{mhchem}
\newcommand{\braket}[3] {{\langle #1 | #2 | #3 \rangle}}
\newcommand{\brket}[2] {{\langle #1 | #2  \rangle}}

\newcommand{\dbd}[2] {{\frac{d #1}{d #2}}}

\newcommand{\figref}[1]{Figure \ref{#1}}
\def\zb{\bar{z}}
\def\bfr{{\bf r}}
\begin{document}
\title[Holomorphic HF and CI]{Holomorphic Hartree--Fock Theory and Configuration Interaction}
\author{Hamish~G.~Hiscock}
\author{Alex~J.~W.~Thom}
\email{ajwt3@cam.ac.uk}
\affiliation{University Chemical Laboratory, Lensfield Road, Cambridge CB2 1EW, United Kingdom}
\date{\today}
\begin{abstract}
We investigate the Hartree-Fock solutions to \ce{H2} in a minimal basis.
  We note the properties of the solutions and their disappearance with geometry and propose a new method, Holomorphic Hartree--Fock theory, where we modify the SCF equations to avoid disappearance of the solutions.
We use these solutions as a basis for a non-orthogonal Configuration Interaction to produce a smooth binding curve over a complete range of geometries.
\end{abstract}
\maketitle
\section{Introduction}
Self-Consistent Field (SCF) electronic structure methods, which include Hartree--Fock (HF) and Density Functional Approximations (DFAs) are presently the bedrock of Quantum Chemistry, whether they be used in their own right or as a foundation for more accurate correlation treatments.
In essence, applying an SCF method equates to minimising an energy functional with respect to varying a set of orbitals within a given basis.
This is often recast as an iterative diagonalization procedure, but may also be regarded as a (matrix) polynomial whose roots are to be found, and this procedure is equivalent to locating the stationary points of the energy with respect to non-trivial changes of the orbitals.
It has long been known that, owing to this non-linear form, the SCF equations admit to many solutions, and recently we and others have been interested in both finding the solutions\cite{ThomMetadynamics} as well as in their physical meaning\cite{BesleyGill_09JCP,GilbertGill_08JPCA,ErshovaBesley_11CPL,HansonHeineBesley_13JCP} and other uses\cite{NOCIMetadynamics,KrausbeckBearpark_14CTC}.

It is somewhat surprising to us that, except for some relatively unknown theoretical studies\cite{Fukutome_71PTP,Fukutome_81IJQC}, the nature, number, and existence properties of the SCF solutions are basically unknown, especially given that they are so fundamental in quantum chemistry.
A typical SCF calculation might involve constructing a guess density and then iterating the SCF until convergence.  The more fastidious computational chemist will do a stability analysis\cite{Thouless_60NP,SeegerPople_77JCP} to ensure that such solutions are indeed local minima, continuing downwards in energy until a local minimum is found.
Despite a range of SCF convergence methods commonly available\cite{PulayDIIS,EDIIS,RoothaanHall,ARH}, none guarantee that a global minimum is found, and so the only effective approach to ensuring that a given solution is the lowest energy is by some form of random searching\cite{deAndradeMalbouisson_05IJQC,deAndradeMalbouisson_06IJQC,MalbouissondeAndrade_12AM} which is very seldom done (or at least documented in the literature).
In response to Thom and Head-Gordon's SCF Metadynamics work on locating SCF solutions\cite{ThomMetadynamics}, Li and Paldus produced a careful series of papers\cite{LiPaldus_09JCP,LiPaldus_09IJQC,LiPaldus_09PCCP} investigating the broken symmetry solutions of homonuclear diatomics and ABA triatomics using Thouless stability anaylsis\cite{Thouless_60NP}, and more recently there have been a number of investigations of broken symmetry solutions\cite{SundstromHeadGordon_14JCP,CuiScuseria_13JCP,JimenezHoyosScuseria_11JCTC} and restoration of symmetry\cite{JimenezHoyosScuseria_12JCP}.
Despite this work there is little acknowledgement of the existence of many SCF solutions in the wider computational community.

\begin{figure}
\includegraphics[scale=0.55]{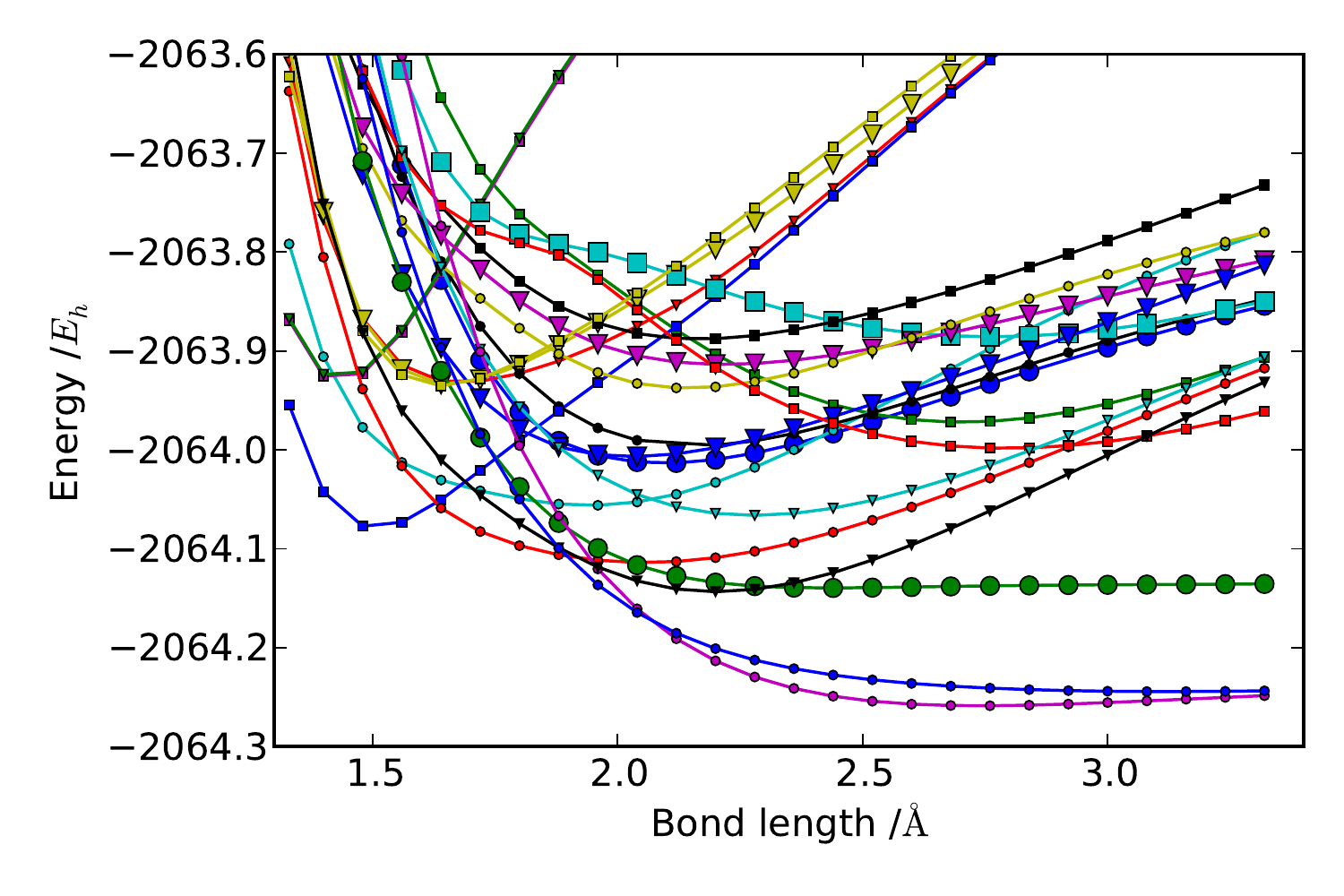}
\caption{Singlet RHF solutions for \ce{Cr2} in the STO-3G basis\cite{PietroHehre_83JCC}. Solutions were located via metadynamics and curve-following.  Curves with small markers are multiply degenerate and large markers singly degenerate.  There is no guarantee that this list of states is exhaustive.}%
\label{fig-Cr2}
\end{figure}

\begin{figure}
\includegraphics[scale=0.55]{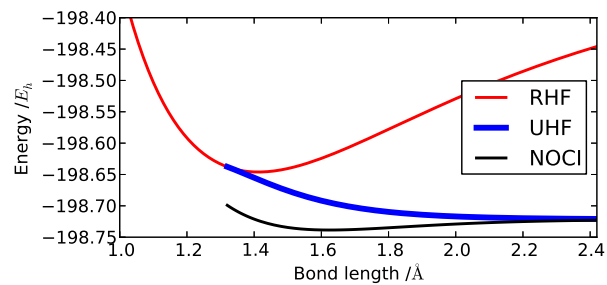}
\caption{RHF and UHF (doubly degenerate) solutions of \ce{F2} in a cc-pVDZ basis\cite{Dunning_89JCP}.  The black line gives the lowest solution when these are used together as a basis for NOCI.  Where only the RHF solution exists, the NOCI solution will correspond to that.}
\label{fig-F2}
\end{figure}
 Thus the most common approach to this multitude of solutions appears to be to mostly ignore them.
In transition metal compounds, for example, we have certainly found solutions\cite{LOBA} which are lower in energy than the local minima found from common guesses, and we suspect this to be a widespread problem.
\figref{fig-Cr2} shows, for example, a number of RHF states of the chromium dimer in a STO-3G basis which cross at different geometries, with four different states having the lowest energy at some point along the binding curve.

Whilst the existence of many SCF solutions might just appear an irksome problem in SCF theory, a number of authors\cite{BesleyGill_09JCP,GilbertGill_08JPCA,ErshovaBesley_11CPL,HansonHeineBesley_13JCP} have sought to interpret the higher energy solutions as corresponding to excited states of the system, and, when present, the SCF solutions do indeed appear to correspond to physical states of the system.
Motivated by this, one of us has shown that in some systems the SCF solutions can be interpreted as quasidiabatic states, and used them as a basis for (non-orthogonal) Configuration Interaction calculations\cite{Malmqvist_86IJQC,NOCIMetadynamics,SundstromHeadGordon_14JCP}, where they reproduce avoided crossing and conical intersections.

Such calculations are, however, more generally thwarted by coalescence and disappearance of solutions (at for example the Coulson-Fischer point\cite{CoulsonFischer}), causing discontinuities in binding curves.
  \figref{fig-F2} shows the Non-Orthogonal CI of the RHF and UHF solutions for the fluorine molecule which are extremely plausible until the Coulson-Fischer point.
Understanding the disappearance of these solutions is therefore crucial if they are to be used as such or interpreted as anything other than artefacts. 

In this paper we go back to basics and thoroughly investigate the solutions to one of the simplest chemical systems, \ce{H2}, in a minimal basis.
  We note the properties of the solutions and their disappearance with geometry and propose a new method, Holomorphic Hartree--Fock theory, where we modify the SCF equations to avoid disappearance of the solutions.
With a fixed number of solutions across the whole binding curve we conclude by showing that these new solutions can be used as a basis for a non-orthogonal Configuration Interaction producing smooth binding curves over a complete range of geometries.
\section{Computational Details}
SCF calculations were performed in a modified version of Q-Chem 4.0\cite{QChem} with additional processing using SymPy\cite{SymPy} and SciPy\cite{SciPy} and figures plotted with matplotlib\cite{Hunter:2007}.

\section{SCF Equations}
Beginning with the single-particle basis set in 3-dimensional space, denoted $\chi_\mu(\bfr)$ and generally constructed from atom-centred functions, we may construct an orthonormal basis, $\tilde\chi_\mu$, which spans the same space, and express other functions, such as molecular orbitals, as an expansion in this basis,
\begin{equation}
\phi_i=\sum_\mu \tilde\chi_\mu C^\mu_{\cdot i},
\end{equation}
where we are using the tensor notation of Head-Gordon \textit{et al.}\cite{HeadGordonWhite_98JCP}, and will use the Einstein summation convention for repeated indices when no explicit summation is specified.  Here the coefficients $C^\mu_{\cdot i}$ may be complex, and, as the basis $\chi_\mu$ increases towards the complete, any complex-valued function in the Hilbert space may be expanded in this form.

We will primarily be concerned in this paper with the Hartree--Fock Self-Consistent Field Approximation, the algorithm and equations for which can be formulated thus:
\begin{enumerate}
\item Begin with a guess for coefficients $C^\mu_{\cdot i}$.
\item Form the one-particle density matrix $P^{\mu\nu}=\sum_i^N C^\mu_{\cdot i} C^{*\mu}_i$.

\item The energy is formed as a functional of density $E(P^{\mu\nu})$,
\begin{equation}
E=h_{\mu\nu}P^{\nu\mu}+\frac{1}{2}P^{\mu\sigma}\mathbb{I}_{\mu\nu\sigma\tau}P^{\nu\tau}
\end{equation}
where the one-electron integrals are defined as $h_{\mu\nu}=\braket{\tilde\chi_\mu}{\hat{h}}{\tilde\chi_\nu}$ for $\hat{h}$ containing kinetic energy and external potential operators, and the two-electron antisymmetrized Coulomb integrals are $\mathbb{I}_{\mu\nu\sigma\tau}=\brket{\mu\nu}{\sigma\tau}-\brket{\mu\nu}{\tau\sigma}$
for
\begin{equation}
\brket{\mu\nu}{\sigma\tau}=\iint\frac{\tilde\chi_\mu^*(\bfr_1)\tilde\chi^*_\nu(\bfr_2)\tilde\chi_\sigma(\bfr_1)\tilde\chi_\tau(\bfr_2)}{|\bfr_1-\bfr_2|}d^3\bfr_1d^3\bfr_2
\end{equation}
\item We solve for $\dbd{E}{P^{\mu\nu}}=0$ (keeping number of electrons fixed) commonly leading to an iterative set of diagonalizations.
\end{enumerate}
Though not normally written as such, the equations can equally well be written as a function of the coefficients $C^\mu_{\cdot i}$ by including orthogonality of the orbitals in a Lagrangian
\begin{equation}
\Lambda=E[C^\mu_{\cdot i}] - \sum_{ij}\lambda_{ij}(C^{*\mu}_iC^\mu_{\cdot j} -\delta_{ij})
\end{equation}
and solving $\dbd{\Lambda}{C^\mu_{\cdot i}}=0$, resulting in a coupled set of polynomials in coefficients $C$.
It is this formulation we would like to consider.
\section{SCF Equations for \ce{H2}}
\begin{figure}
\hspace{-3em}\includegraphics[scale=0.50]{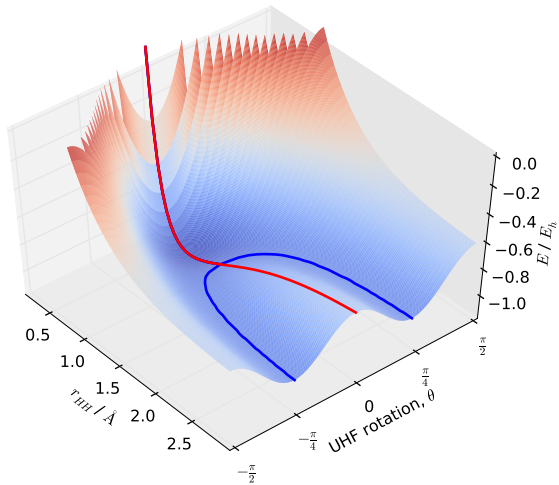}
\caption{The energy surface for \ce{H2}. UHF rotation indicates the angle $\theta$ by which the $\alpha$ MO has mixed together the $\sigma_g$ and $\sigma_u$ orbitals as specified in the text.  The red curve shows the RHF solution as a local minimum against rotation for $r_{HH}<1.2$\AA, becoming a maximum after this; the blue curves are the two degenerate UHF solutions.}
\label{fig-H2EvsRot}
\end{figure}
\begin{figure}
\includegraphics[scale=0.40]{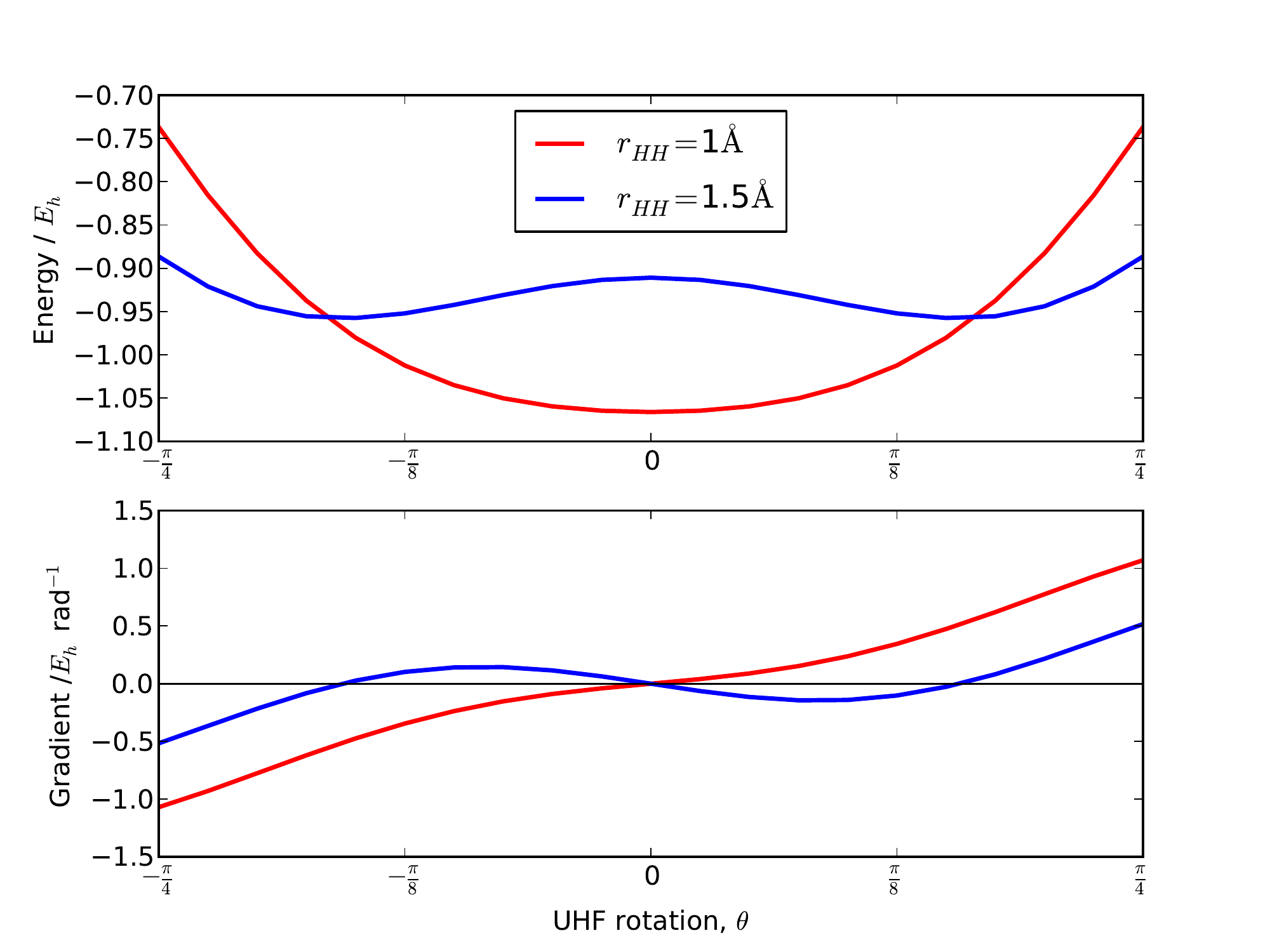}
\caption{Two slices through the energy surface for \ce{H2}, before (red, $r_{HH}=1\,$\AA) and after (blue, $r_{HH}=1.5\,$\AA) the Coulson-Fischer point, showing the energy (top) and gradient with respect to orbital rotation (bottom), showing one (red) and three (blue) stationary points.}
\label{fig-H2slices}
\end{figure}
\begin{figure*}
\hspace{-5em}\includegraphics[scale=0.50]{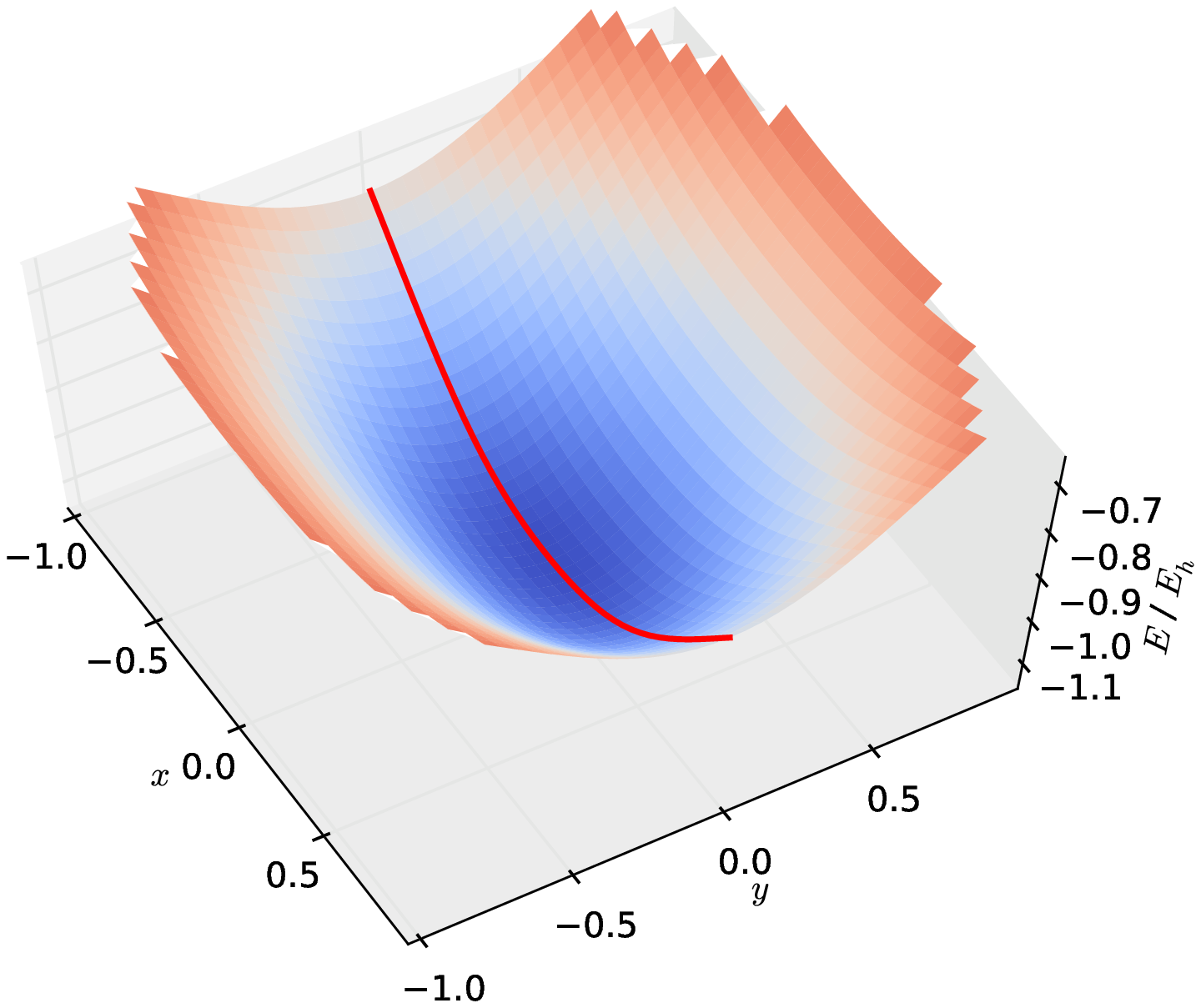}
\includegraphics[scale=0.50]{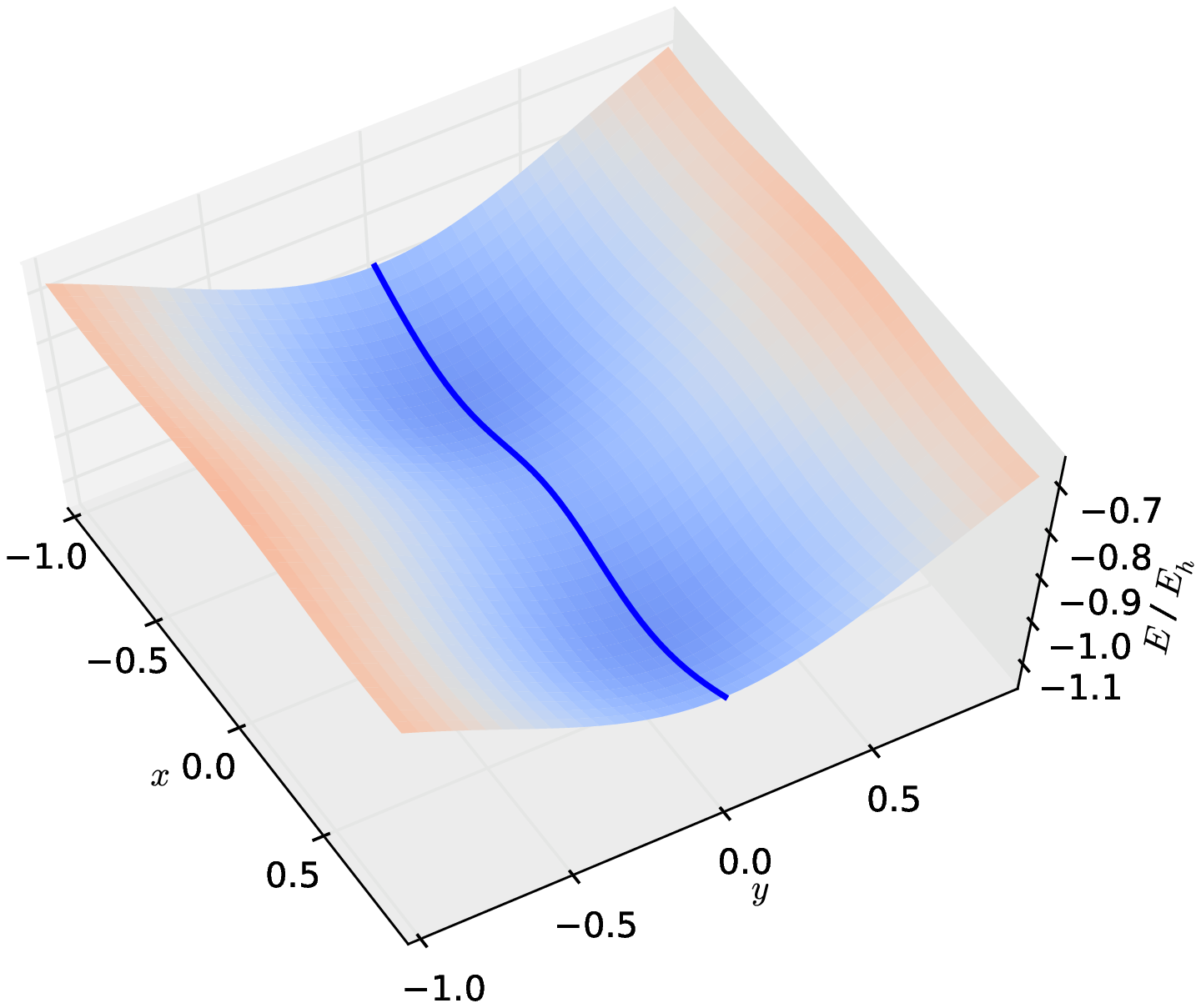}
\caption{The energy surface for complex $z=x+iy$ for (left) $r_{HH}=1\,$\AA\ and $r_{HH}=1.5\,$\AA.  There are no additional minima at complex $z$.  The coloured curves correspond to those in \figref{fig-H2slices}}
\label{fig-H2Complex}
\end{figure*}
For the very simple case of \ce{H2} in a minimal basis (here we choose STO-3G \cite{HehrePople_69JCP}) with a single atomic orbital sited on each atom, it is well-known that as the bond length increases past the Coulson-Fischer point, the restricted Hartree-Fock solution (where both $\alpha$ and $\beta$ spin molecular orbitals have the same spatial form) becomes unstable with respect to a symmetry-broken Unrestricted Hartree-Fock solution where the $\alpha$ and $\beta$ molecular orbitals move to becoming localized on separate atoms.  A convenient representation for these solutions is to consider the two orbitals as consisting of rotations of the symmetry orbitals, $\sigma_g$ and $\sigma_u$:
\begin{eqnarray}
\phi_\alpha=\sigma_g\cos\theta + \sigma_u\sin\theta, \\
\phi_\beta=\sigma_g\cos\theta -  \sigma_u\sin\theta.
\end{eqnarray}
This is shown in \ref{fig-H2EvsRot}.
Here, there is a single electronic degree of freedom, parameterized by $\theta$, and for a given geometry, we may plot energy against this parameter and find the solutions to the SCF equations being the stationary points of these functions.  Two such curves are given in \figref{fig-H2slices}.
Given the form of the curves, it is tempting to see these as quartic polynomials in $\theta$  with one and real three roots.
Recalling the Fundamental Theorem of Algebra, which states that every non-zero, single-variable, degree-$n$ polynomial with complex coefficients has, counted with multiplicity, exactly $n$ roots, we speculated that the missing roots might correspond to orbitals with complex coefficients.
To investigate this further it is convenient to transform to a different parameterization and we choose to write the orbitals in terms of (complex) parameter $z$,
\begin{eqnarray}
\phi_\alpha=\frac1{\sqrt{1+|z|^2}}\sigma_g + \frac{z}{\sqrt{1+|z|^2}}\sigma_u \\
\phi_\beta=\frac1{\sqrt{1+|z|^2}}\sigma_g -  \frac{z}{\sqrt{1+|z|^2}}\sigma_u,
\end{eqnarray}
where $z$ can be equated with $\tan\theta$.  For this system, taking advantage of the symmetry of the integrals, the energy becomes a function of $z$,
\begin{widetext}
$$
E(z)=\frac2{1+z\zb}(h_{gg}+z\zb h_{uu}) + \frac1{(1+z\zb)^2}\left(\brket{gg}{gg}-(z^2+\zb^2)\brket{gg}{uu}+(z\zb)^2\brket{uu}{uu} + 2z\zb\brket{gu}{gu}-2z\zb\brket{gu}{ug}\right),
$$
\end{widetext}
where $\zb$ is the complex conjugate of $z$.
As $E(z)$ is purely real, we might hope that we can locate solutions $\dbd{E(z)}{z}=0$, where $z$ is complex; since $\zb$ is simply a function of $z$, it is not independent and need not be explicitly considered.
Indeed as $E(z)$ is a strictly real function, using $z=x+iy$ it can be viewed as a surface $E(x,y)$ which has stationary points which can be located by standard methods.

A plot of the energy surfaces for $r_{HH}=1\,$\AA{} and $r_{HH}=1.5\,$\AA{} are given in \figref{fig-H2Complex}.
While this view is appealing, as can be seen in the figure, it unfortunately does not lead to finding any additional complex solutions. 
To understand why, we must consider at $E$ as a complex function.
\section{Holomorphic Functions}
\begin{figure*}
\hspace{-5em}\includegraphics[scale=0.50]{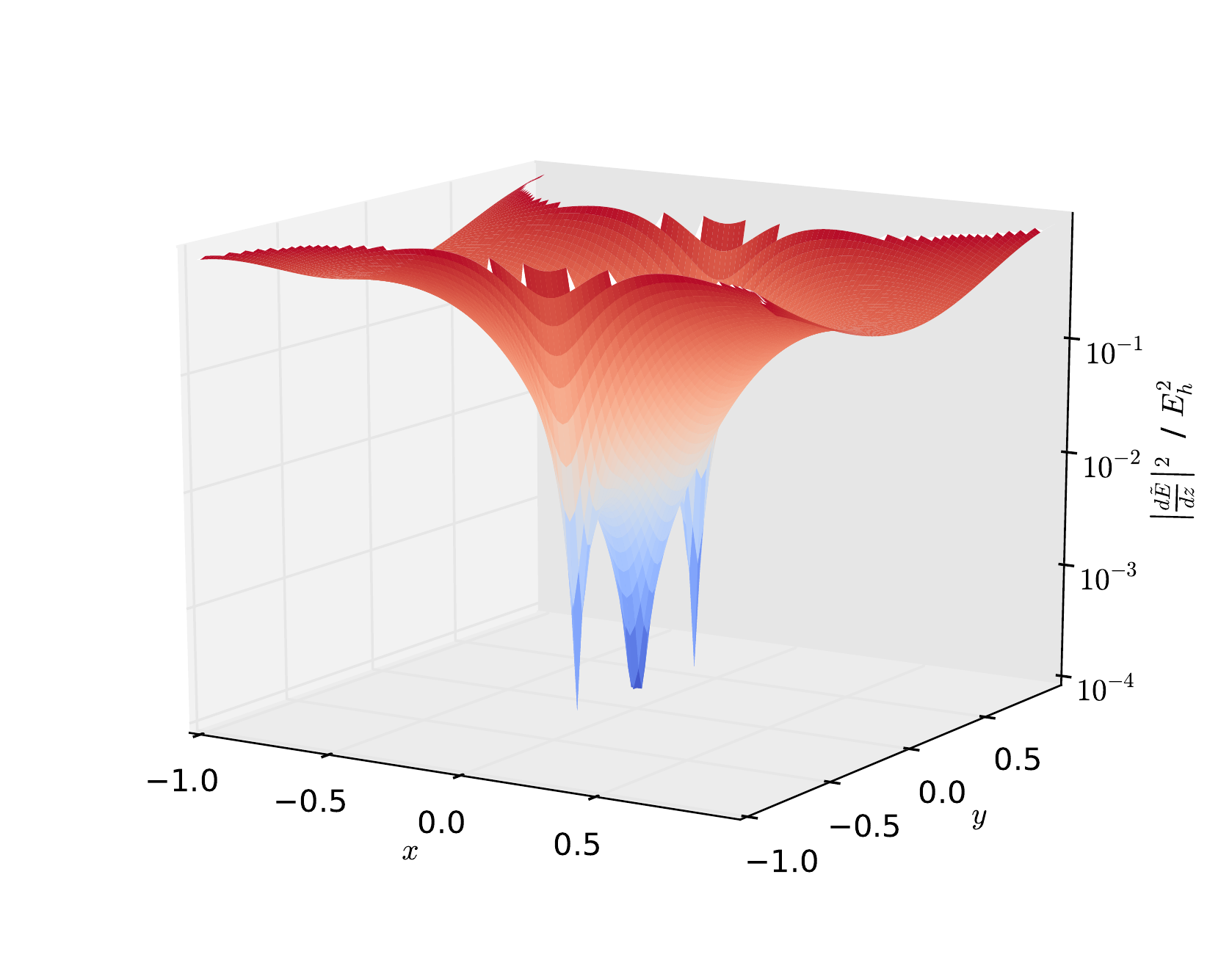}
\includegraphics[scale=0.50]{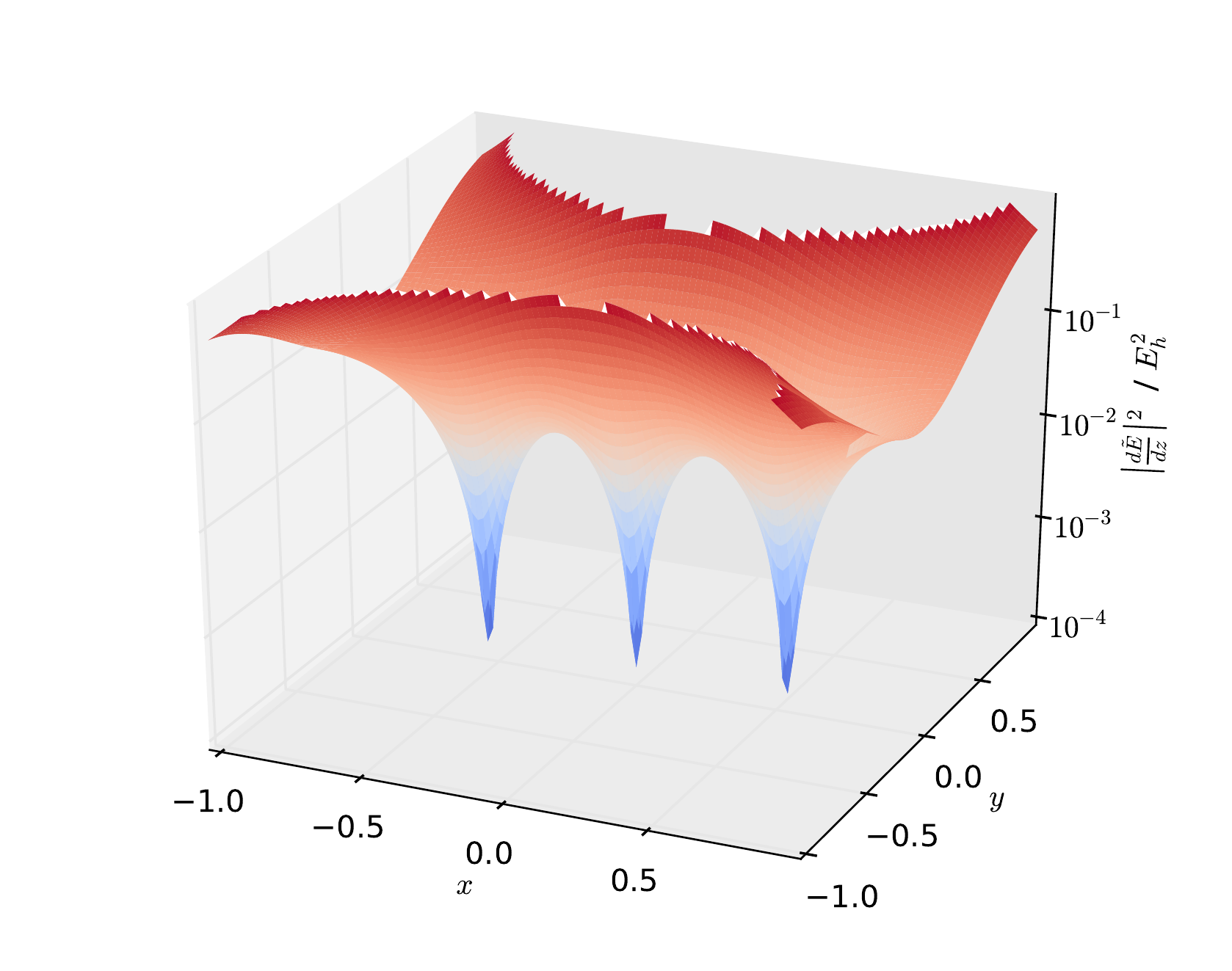}
\caption{The square magnitude of the gradient of the holomorphized energy $\tilde E(z)$ plotted in the complex plane $z=x+iy$.  The left picture shows $r_{HH}=1\,$\AA{} with a root at $z=0$ and two additional roots at $z=\pm0.361110i$; the right shows $r_{HH}=1.5\,$\AA{} with the conventional solutions, $z=0, \pm0.536989$.}
\label{fig-H2HoloEGrad}
\end{figure*}
Despite the energy $E(z)$ being a real function of $z$ by construction, in general, functions of complex variables are usually complex valued, and such functions and their derivatives are the subject of the theory of complex analysis.
While a full exposition of the field is beyond the scope of this paper, there are a number of useful results it can bring to bear.
Primarily, it should be noted that not all functions of complex variables have well-defined complex derivatives, and only a subset of functions which obey the Cauchy-Riemann conditions are differentiable.

The most relevant form of these conditions in this case is the following:  A function $f(z)$ of complex variable $z$ is complex-differentiable if it has no dependence on $\zb$.
Such a complex-differentiable function is known as \textit{holomorphic}.
Immediately we can see that the energy function $E(z)$ does not satisfy these conditions.
Unfortunately, the Fundamental Theorem of Algebra which guarantees that complex solutions will exist is only
valid for holomorphic polynomials (which do not contain any explicit $\zb$ dependence), and this is why we have not been able to locate further complex solutions for this energy function.

This problem immediately presents its own solution, however, in that if we wish to rely on the Fundamental Theorem of Algebra, we must convert the energy expression into one which is not dependent on $\zb$.
We may do this by creating $\tilde E(z)$ from $E(z)$ by replacing all instances of $\zb$ with $z$ and call this process \textit{holomorphizing}.
A consequence of this is that $\tilde E(z)$ is no longer a real-valued function, though we note that where $z$ is real (i.e. for all stationary points of $E$), $\tilde E$ will take the same value as $E$ and because of complex conjugation symmetry will still be stationary in the imaginary direction, so no stationary points of the original $E$ have been lost.

Being complex-valued, the $\tilde E$ surface is extremely difficult to visualize, so we plot instead the square magnitude of the derivative of the surface in \figref{fig-H2HoloEGrad}.
  Both sides of the the Coulson-Fischer point there are the same number of zeroes of gradient (we denote these as holomorphic UHF (hUHF) solutions), with the additional ones at $r_{HH}=1\,$\AA{} having complex orbitals.

\begin{figure*}
\includegraphics[scale=0.45]{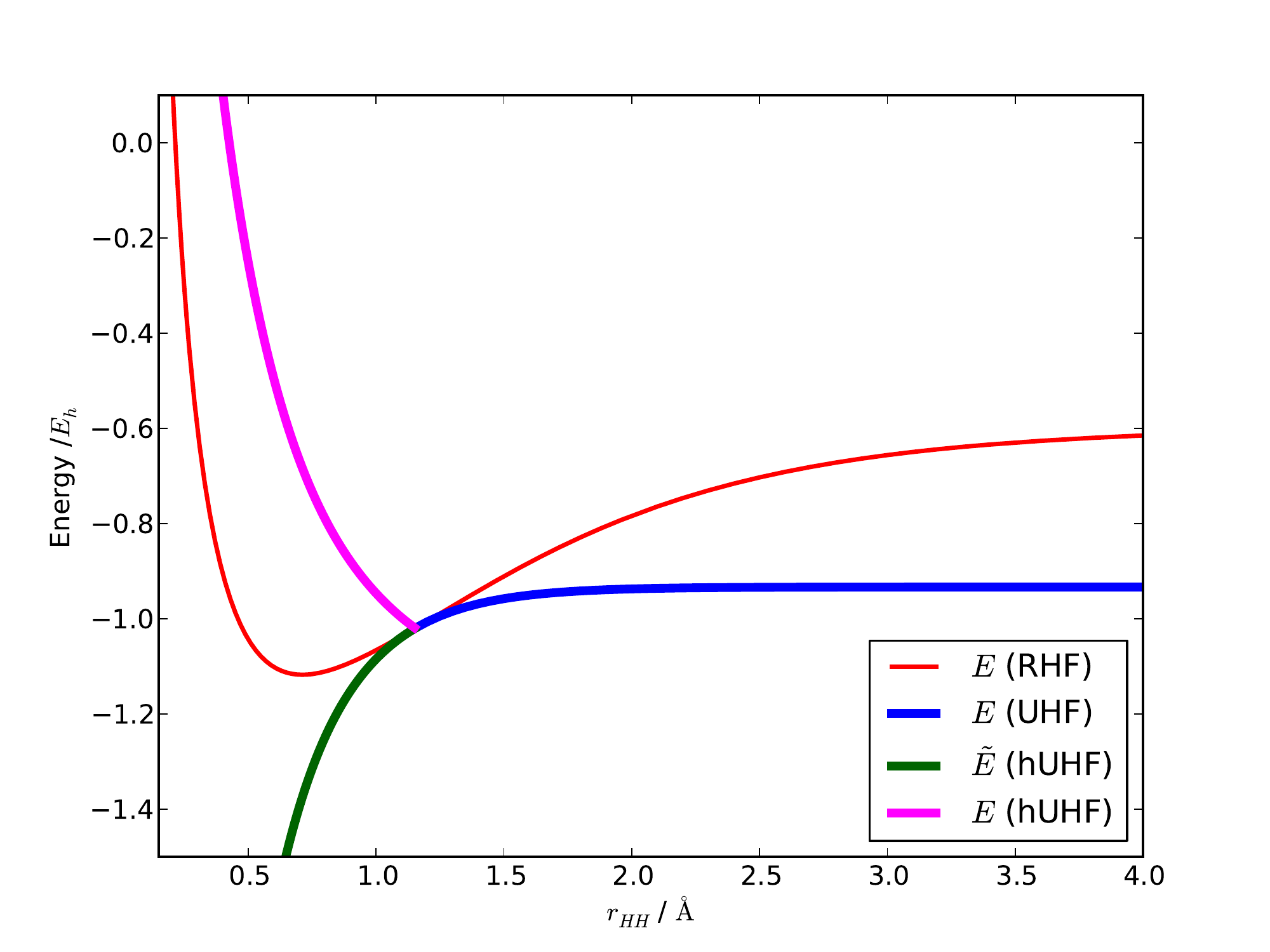}
\hspace{-2em}\includegraphics[scale=0.45]{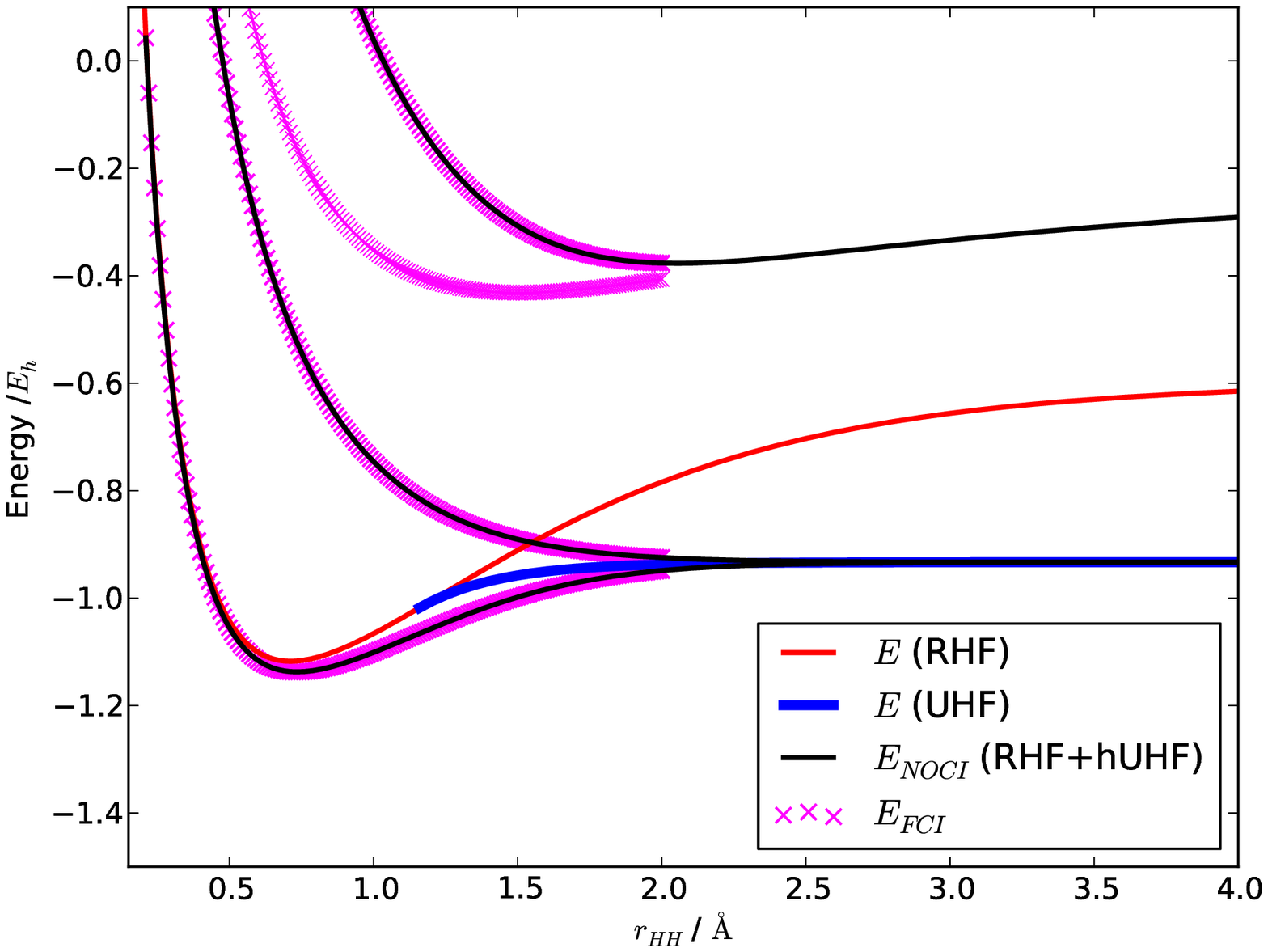}
\caption{Left: a comparison of energies for \ce{H2} for RHF, UHF and holomorphic UHF (hUHF) stationary points.  For hUHF, the holomorphic energy $\tilde{E}$ and the real energy $E$ are plotted for the same orbitals. For $r_{HH}>1.15$\AA, the hUHF are identical to the UHF solutions; right: Energies from a Non-Orthogonal Configuration Interaction of the RHF and hUHF solutions which are identical to a subset of the $M_S=0$ Full Configuration Interaction energies in this basis.  The symmetries of the FCI states are (from the bottom) $^1\Sigma_g^+, ^3\!\Sigma_u^+, ^1\!\Sigma_u^+, ^1\!\Sigma_g^+$.}
\label{fig-H2holo}
\end{figure*}

For this system it is trivial to locate these hUHF solutions across the whole range of $r_{HH}$ these are plotted in \figref{fig-H2holo}.
It can be seen that the holomorphic energies $\tilde{E}$ are manifestly not variational, and at small bond lengths are lower than the RHF energies.  We note that the hUHF $\tilde E$ do not diverge to negative infinity, but have a minimum at about 0.2\AA{} and then increase.
With the orbitals corresponding to the hUHF solutions, the real energy $E$ can be calculated, and it is found to vary continuously, and (as is required) to be higher than the RHF energy before the Coulson-Fischer point.
After the Coulson-Fischer point, the hUHF solutions correspond exactly to the UHF solutions, and so provide a set of solutions which exist at all geometries.
\section{Non-Orthogonal CI}
The motivation for this study of HF solutions arose in using them in a Non-Orthogonal Configuration Interaction \cite{NOCIMetadynamics} and the difficulties caused by such solutions disappearing.
By solving the hUHF equations, we have a modified theory which has a constant number of solutions as geometry changes, and these are eminently amenable as a basis for describing molecular dissociation.
Using the three hUHF solutions as a basis for Configuration Interaction, we can construct the Hamiltonian and overlap matrices in this basis (being $3\times3$ matrices), and solve the generalized eigenvalue problem to recover energies.
We refer the reader to reference \citenum{NOCIMetadynamics} for details as to how to perform this calculation.

The results are plotted in the right hand side of \figref{fig-H2holo}, and compared to the Full Configuration Interaction in this basis (with the restriction that $M_S=0$).
We find that the hUHF-NOCI solutions provide identical, and most importantly, smooth curves for three of the FCI states%
\footnote{We note that in this small example set comprising the two hUHF and single RHF solutions happen to span the  subspace from the two $^1\Sigma_g^+$ and the $M_S=0$ $^3\!\Sigma_u^+$ solution which results in this smoothness.
This spanning is by no means required of the formalism, however, and a simple thought experiment performed with the addition of an extra determinant (for example constructed from two $p$-orbitals) which couples to the FCI space would change the FCI energies, but not the hUHF-NOCI energies, and so the smoothness would be maintained in this larger space.}.
Including any of the higher remaining UHF or RHF states in the NOCI recovers the remaining FCI solution.
\section{Conclusion}
We have demonstrated that the Hartree-Fock SCF equations can be modified so as to be holomorphic, and therefore admit to a constant number of solutions across geometries.
Where RHF and UHF solutions exist, the holomorphic solutions are identical, but at geometries when conventional SCF solutions disappear, complex holomorphic solutions appear.
These solutions are stationary values of the non-variational holomorphic energy, but have real energy expectation values
which are above the lowest RHF solution.
Crucially, these solutions can be used as a basis for Non-Orthogonal Configuration Interaction and have been shown to
provide smooth energy curves.

In future work we hope to extend this formalism to larger systems using more conventional solution techniques, and show this to be a usable method for locating SCF states.

AJWT thanks the Royal Society for a University Research Fellowship and summer project funding for HGH, and Dr James Spencer and Dr George Booth for their helpful comments on the manuscript.
\bibliography{SCFHolo}

\begin{thebibliography}{36}
\expandafter\ifx\csname natexlab\endcsname\relax\def\natexlab#1{#1}\fi
\expandafter\ifx\csname bibnamefont\endcsname\relax
  \def\bibnamefont#1{#1}\fi
\expandafter\ifx\csname bibfnamefont\endcsname\relax
  \def\bibfnamefont#1{#1}\fi
\expandafter\ifx\csname citenamefont\endcsname\relax
  \def\citenamefont#1{#1}\fi
\expandafter\ifx\csname url\endcsname\relax
  \def\url#1{\texttt{#1}}\fi
\expandafter\ifx\csname urlprefix\endcsname\relax\def\urlprefix{URL }\fi
\providecommand{\bibinfo}[2]{#2}
\providecommand{\eprint}[2][]{\url{#2}}

\bibitem[{\citenamefont{Thom and Head-Gordon}(2008)}]{ThomMetadynamics}
\bibinfo{author}{\bibfnamefont{A.~J.~W.} \bibnamefont{Thom}} \bibnamefont{and}
  \bibinfo{author}{\bibfnamefont{M.}~\bibnamefont{Head-Gordon}},
  \bibinfo{journal}{Phys. Rev. Lett.} \textbf{\bibinfo{volume}{101}},
  \bibinfo{pages}{193001} (\bibinfo{year}{2008}).

\bibitem[{\citenamefont{Besley et~al.}(2009)\citenamefont{Besley, Gilbert, and
  Gill}}]{BesleyGill_09JCP}
\bibinfo{author}{\bibfnamefont{N.~A.} \bibnamefont{Besley}},
  \bibinfo{author}{\bibfnamefont{A.~T.~B.} \bibnamefont{Gilbert}},
  \bibnamefont{and} \bibinfo{author}{\bibfnamefont{P.~M.~W.}
  \bibnamefont{Gill}}, \bibinfo{journal}{J. Chem. Phys.}
  \textbf{\bibinfo{volume}{130}}, \bibinfo{pages}{124308}
  (\bibinfo{year}{2009}).

\bibitem[{\citenamefont{Gilbert et~al.}(2008)\citenamefont{Gilbert, Besley, and
  Gill}}]{GilbertGill_08JPCA}
\bibinfo{author}{\bibfnamefont{A.~T.~B.} \bibnamefont{Gilbert}},
  \bibinfo{author}{\bibfnamefont{N.~A.} \bibnamefont{Besley}},
  \bibnamefont{and} \bibinfo{author}{\bibfnamefont{P.~M.~W.}
  \bibnamefont{Gill}}, \bibinfo{journal}{J. Phys. Chem. A}
  \textbf{\bibinfo{volume}{112}}, \bibinfo{pages}{13164}
  (\bibinfo{year}{2008}).

\bibitem[{\citenamefont{Ershova and Besley}(2011)}]{ErshovaBesley_11CPL}
\bibinfo{author}{\bibfnamefont{O.~V.} \bibnamefont{Ershova}} \bibnamefont{and}
  \bibinfo{author}{\bibfnamefont{N.~A.} \bibnamefont{Besley}},
  \bibinfo{journal}{Chem. Phys. Lett.} \textbf{\bibinfo{volume}{513}},
  \bibinfo{pages}{179} (\bibinfo{year}{2011}).

\bibitem[{\citenamefont{Hanson-Heine et~al.}(2013)\citenamefont{Hanson-Heine,
  George, and Besley}}]{HansonHeineBesley_13JCP}
\bibinfo{author}{\bibfnamefont{M.~W.~D.} \bibnamefont{Hanson-Heine}},
  \bibinfo{author}{\bibfnamefont{M.~W.} \bibnamefont{George}},
  \bibnamefont{and} \bibinfo{author}{\bibfnamefont{N.~A.}
  \bibnamefont{Besley}}, \bibinfo{journal}{J. Chem. Phys.}
  \textbf{\bibinfo{volume}{138}}, \bibinfo{pages}{064101}
  (\bibinfo{year}{2013}).

\bibitem[{\citenamefont{Thom and Head-Gordon}(2009)}]{NOCIMetadynamics}
\bibinfo{author}{\bibfnamefont{A.~J.~W.} \bibnamefont{Thom}} \bibnamefont{and}
  \bibinfo{author}{\bibfnamefont{M.}~\bibnamefont{Head-Gordon}},
  \bibinfo{journal}{J. Chem. Phys.} \textbf{\bibinfo{volume}{131}},
  \bibinfo{pages}{123114} (\bibinfo{year}{2009}).

\bibitem[{\citenamefont{Krausbeck et~al.}(2014)\citenamefont{Krausbeck,
  Mendive-Tapia, Thom, and Bearpark}}]{KrausbeckBearpark_14CTC}
\bibinfo{author}{\bibfnamefont{F.}~\bibnamefont{Krausbeck}},
  \bibinfo{author}{\bibfnamefont{D.}~\bibnamefont{Mendive-Tapia}},
  \bibinfo{author}{\bibfnamefont{A.~J.~W.} \bibnamefont{Thom}},
  \bibnamefont{and} \bibinfo{author}{\bibfnamefont{M.~J.}
  \bibnamefont{Bearpark}}, \bibinfo{journal}{Comp. Theor. Chem.}
  (\bibinfo{year}{2014}).

\bibitem[{\citenamefont{Fukutome}(1971)}]{Fukutome_71PTP}
\bibinfo{author}{\bibfnamefont{H.}~\bibnamefont{Fukutome}},
  \bibinfo{journal}{Prog. Theor. Phys.} \textbf{\bibinfo{volume}{45}},
  \bibinfo{pages}{1382} (\bibinfo{year}{1971}).

\bibitem[{\citenamefont{Fukutome}(1981)}]{Fukutome_81IJQC}
\bibinfo{author}{\bibfnamefont{H.}~\bibnamefont{Fukutome}},
  \bibinfo{journal}{Int. J. Quant. Chem.} \textbf{\bibinfo{volume}{20}},
  \bibinfo{pages}{955} (\bibinfo{year}{1981}).

\bibitem[{\citenamefont{Thouless}(1960)}]{Thouless_60NP}
\bibinfo{author}{\bibfnamefont{D.~J.} \bibnamefont{Thouless}},
  \bibinfo{journal}{Nucl. Phys.} \textbf{\bibinfo{volume}{21}},
  \bibinfo{pages}{225} (\bibinfo{year}{1960}).

\bibitem[{\citenamefont{Seeger and Pople}(1977)}]{SeegerPople_77JCP}
\bibinfo{author}{\bibfnamefont{R.}~\bibnamefont{Seeger}} \bibnamefont{and}
  \bibinfo{author}{\bibfnamefont{J.~A.} \bibnamefont{Pople}},
  \bibinfo{journal}{J. Chem. Phys.} \textbf{\bibinfo{volume}{66}},
  \bibinfo{pages}{3045} (\bibinfo{year}{1977}).

\bibitem[{\citenamefont{Pulay}(1980)}]{PulayDIIS}
\bibinfo{author}{\bibfnamefont{P.}~\bibnamefont{Pulay}},
  \bibinfo{journal}{Chem. Phys. Lett.} \textbf{\bibinfo{volume}{73}},
  \bibinfo{pages}{392} (\bibinfo{year}{1980}).

\bibitem[{\citenamefont{Kudin et~al.}(2002)\citenamefont{Kudin, Scuseria, and
  Canc{\`e}s}}]{EDIIS}
\bibinfo{author}{\bibfnamefont{K.~N.} \bibnamefont{Kudin}},
  \bibinfo{author}{\bibfnamefont{G.~E.} \bibnamefont{Scuseria}},
  \bibnamefont{and}
  \bibinfo{author}{\bibfnamefont{E.}~\bibnamefont{Canc{\`e}s}},
  \bibinfo{journal}{J. Chem. Phys.} \textbf{\bibinfo{volume}{116}},
  \bibinfo{pages}{8255} (\bibinfo{year}{2002}).

\bibitem[{\citenamefont{Roothaan}(1951)}]{RoothaanHall}
\bibinfo{author}{\bibfnamefont{C.~C.~J.} \bibnamefont{Roothaan}},
  \bibinfo{journal}{Rev. Mod. Phys.} \textbf{\bibinfo{volume}{23}},
  \bibinfo{pages}{69} (\bibinfo{year}{1951}).

\bibitem[{\citenamefont{{H\o st} et~al.}(2008)\citenamefont{{H\o st}, {Jans\'\i
  k}, Olsen, {J\o rgensen}, Reine, and Helgaker}}]{ARH}
\bibinfo{author}{\bibfnamefont{S.}~\bibnamefont{{H\o st}}},
  \bibinfo{author}{\bibfnamefont{B.}~\bibnamefont{{Jans\'\i k}}},
  \bibinfo{author}{\bibfnamefont{J.}~\bibnamefont{Olsen}},
  \bibinfo{author}{\bibfnamefont{P.}~\bibnamefont{{J\o rgensen}}},
  \bibinfo{author}{\bibfnamefont{S.}~\bibnamefont{Reine}}, \bibnamefont{and}
  \bibinfo{author}{\bibfnamefont{T.}~\bibnamefont{Helgaker}},
  \bibinfo{journal}{Phys. Chem. Chem. Phys.} \textbf{\bibinfo{volume}{10}},
  \bibinfo{pages}{5344} (\bibinfo{year}{2008}).

\bibitem[{\citenamefont{de~Andrade et~al.}(2005)\citenamefont{de~Andrade,
  Mundim, and Malbouisson}}]{deAndradeMalbouisson_05IJQC}
\bibinfo{author}{\bibfnamefont{M.~D.} \bibnamefont{de~Andrade}},
  \bibinfo{author}{\bibfnamefont{K.~C.} \bibnamefont{Mundim}},
  \bibnamefont{and} \bibinfo{author}{\bibfnamefont{L.~A.~C.}
  \bibnamefont{Malbouisson}}, \bibinfo{journal}{Int. J. Quant. Chem.}
  \textbf{\bibinfo{volume}{103}}, \bibinfo{pages}{493} (\bibinfo{year}{2005}).

\bibitem[{\citenamefont{de~Andrarade et~al.}(2006)\citenamefont{de~Andrarade,
  Nascimento, Mundim, and Malbouisson}}]{deAndradeMalbouisson_06IJQC}
\bibinfo{author}{\bibfnamefont{M.~D.} \bibnamefont{de~Andrarade}},
  \bibinfo{author}{\bibfnamefont{M.}~\bibnamefont{Nascimento}},
  \bibinfo{author}{\bibfnamefont{K.}~\bibnamefont{Mundim}}, \bibnamefont{and}
  \bibinfo{author}{\bibfnamefont{L.}~\bibnamefont{Malbouisson}},
  \bibinfo{journal}{Int. J. Quant. Chem.} \textbf{\bibinfo{volume}{106}},
  \bibinfo{pages}{2700} (\bibinfo{year}{2006}).

\bibitem[{\citenamefont{Malbouisson et~al.}(2012)\citenamefont{Malbouisson,
  de~Cerqueira~Sobrinho, Nascimento, and {de
  Andrade}}}]{MalbouissondeAndrade_12AM}
\bibinfo{author}{\bibfnamefont{L.~A.~C.} \bibnamefont{Malbouisson}},
  \bibinfo{author}{\bibfnamefont{A.~M.} \bibnamefont{de~Cerqueira~Sobrinho}},
  \bibinfo{author}{\bibfnamefont{M.~A.~C.} \bibnamefont{Nascimento}},
  \bibnamefont{and} \bibinfo{author}{\bibfnamefont{M.~D.} \bibnamefont{{de
  Andrade}}}, \bibinfo{journal}{Appl. Math.} \textbf{\bibinfo{volume}{3}},
  \bibinfo{pages}{1526} (\bibinfo{year}{2012}).

\bibitem[{\citenamefont{Li and Paldus}(2009{\natexlab{a}})}]{LiPaldus_09JCP}
\bibinfo{author}{\bibfnamefont{X.}~\bibnamefont{Li}} \bibnamefont{and}
  \bibinfo{author}{\bibfnamefont{J.}~\bibnamefont{Paldus}},
  \bibinfo{journal}{J. Chem. Phys.} \textbf{\bibinfo{volume}{130}},
  \bibinfo{pages}{084110} (\bibinfo{year}{2009}{\natexlab{a}}).

\bibitem[{\citenamefont{Li and Paldus}(2009{\natexlab{b}})}]{LiPaldus_09IJQC}
\bibinfo{author}{\bibfnamefont{X.}~\bibnamefont{Li}} \bibnamefont{and}
  \bibinfo{author}{\bibfnamefont{J.}~\bibnamefont{Paldus}},
  \bibinfo{journal}{Int. J. Quant. Chem.} \textbf{\bibinfo{volume}{109}},
  \bibinfo{pages}{1756} (\bibinfo{year}{2009}{\natexlab{b}}).

\bibitem[{\citenamefont{Li and Paldus}(2009{\natexlab{c}})}]{LiPaldus_09PCCP}
\bibinfo{author}{\bibfnamefont{X.}~\bibnamefont{Li}} \bibnamefont{and}
  \bibinfo{author}{\bibfnamefont{J.}~\bibnamefont{Paldus}},
  \bibinfo{journal}{Phys. Chem. Chem. Phys.} \textbf{\bibinfo{volume}{11}},
  \bibinfo{pages}{5281} (\bibinfo{year}{2009}{\natexlab{c}}).

\bibitem[{\citenamefont{Sundstrom and
  Head-Gordon}(2014)}]{SundstromHeadGordon_14JCP}
\bibinfo{author}{\bibfnamefont{E.~J.} \bibnamefont{Sundstrom}}
  \bibnamefont{and}
  \bibinfo{author}{\bibfnamefont{M.}~\bibnamefont{Head-Gordon}},
  \bibinfo{journal}{J. Chem. Phys.} \textbf{\bibinfo{volume}{140}},
  \bibinfo{pages}{114103} (\bibinfo{year}{2014}).

\bibitem[{\citenamefont{Cui et~al.}(2012)\citenamefont{Cui, Bulik,
  {Jim\'enez-Hoyos}, Henderson, and Scuseria}}]{CuiScuseria_13JCP}
\bibinfo{author}{\bibfnamefont{Y.}~\bibnamefont{Cui}},
  \bibinfo{author}{\bibfnamefont{I.~W.} \bibnamefont{Bulik}},
  \bibinfo{author}{\bibfnamefont{C.~A.} \bibnamefont{{Jim\'enez-Hoyos}}},
  \bibinfo{author}{\bibfnamefont{T.~M.} \bibnamefont{Henderson}},
  \bibnamefont{and} \bibinfo{author}{\bibfnamefont{G.~E.}
  \bibnamefont{Scuseria}}, \bibinfo{journal}{J. Chem. Phys.}
  \textbf{\bibinfo{volume}{139}}, \bibinfo{pages}{154107}
  (\bibinfo{year}{2012}).

\bibitem[{\citenamefont{{Jim\'enez-Hoyos}
  et~al.}(2011)\citenamefont{{Jim\'enez-Hoyos}, Henderson, and
  Scuseria}}]{JimenezHoyosScuseria_11JCTC}
\bibinfo{author}{\bibfnamefont{C.~A.} \bibnamefont{{Jim\'enez-Hoyos}}},
  \bibinfo{author}{\bibfnamefont{T.~M.} \bibnamefont{Henderson}},
  \bibnamefont{and} \bibinfo{author}{\bibfnamefont{G.~E.}
  \bibnamefont{Scuseria}}, \bibinfo{journal}{J. Comput. Theor. Chem.}
  \textbf{\bibinfo{volume}{7}}, \bibinfo{pages}{2667} (\bibinfo{year}{2011}).

\bibitem[{\citenamefont{{Jim\'enez-Hoyos}
  et~al.}(2012)\citenamefont{{Jim\'enez-Hoyos}, Henderson, Tsuchimoci, and
  Scuseria}}]{JimenezHoyosScuseria_12JCP}
\bibinfo{author}{\bibfnamefont{C.~A.} \bibnamefont{{Jim\'enez-Hoyos}}},
  \bibinfo{author}{\bibfnamefont{T.~M.} \bibnamefont{Henderson}},
  \bibinfo{author}{\bibfnamefont{T.}~\bibnamefont{Tsuchimoci}},
  \bibnamefont{and} \bibinfo{author}{\bibfnamefont{G.~E.}
  \bibnamefont{Scuseria}}, \bibinfo{journal}{J. Chem. Phys.}
  \textbf{\bibinfo{volume}{136}}, \bibinfo{pages}{164109}
  (\bibinfo{year}{2012}).

\bibitem[{\citenamefont{Pietro and Hehre}(1983)}]{PietroHehre_83JCC}
\bibinfo{author}{\bibfnamefont{W.~J.} \bibnamefont{Pietro}} \bibnamefont{and}
  \bibinfo{author}{\bibfnamefont{W.~J.} \bibnamefont{Hehre}},
  \bibinfo{journal}{J. Comp. Chem.} \textbf{\bibinfo{volume}{4}},
  \bibinfo{pages}{241} (\bibinfo{year}{1983}).

\bibitem[{\citenamefont{Dunning}(1989)}]{Dunning_89JCP}
\bibinfo{author}{\bibfnamefont{T.~H.} \bibnamefont{Dunning},
  \bibfnamefont{Jr.}}, \bibinfo{journal}{J. Chem. Phys.}
  \textbf{\bibinfo{volume}{90}}, \bibinfo{pages}{1007} (\bibinfo{year}{1989}).

\bibitem[{\citenamefont{Thom et~al.}(2009)\citenamefont{Thom, Sundstrom, and
  Head-Gordon}}]{LOBA}
\bibinfo{author}{\bibfnamefont{A.~J.~W.} \bibnamefont{Thom}},
  \bibinfo{author}{\bibfnamefont{E.~J.} \bibnamefont{Sundstrom}},
  \bibnamefont{and}
  \bibinfo{author}{\bibfnamefont{M.}~\bibnamefont{Head-Gordon}},
  \bibinfo{journal}{Phys. Chem. Chem. Phys.} \textbf{\bibinfo{volume}{11}},
  \bibinfo{pages}{11297} (\bibinfo{year}{2009}).

\bibitem[{\citenamefont{Malmqvist}(1986)}]{Malmqvist_86IJQC}
\bibinfo{author}{\bibfnamefont{P.~A.} \bibnamefont{Malmqvist}},
  \bibinfo{journal}{Int. J. Quant. Chem.} \textbf{\bibinfo{volume}{30}},
  \bibinfo{pages}{479} (\bibinfo{year}{1986}).

\bibitem[{\citenamefont{Coulson and Fischer}(1949)}]{CoulsonFischer}
\bibinfo{author}{\bibfnamefont{C.~A.} \bibnamefont{Coulson}} \bibnamefont{and}
  \bibinfo{author}{\bibfnamefont{I.}~\bibnamefont{Fischer}},
  \bibinfo{journal}{Philos. Mag.} \textbf{\bibinfo{volume}{40}},
  \bibinfo{pages}{386} (\bibinfo{year}{1949}).

\bibitem[{\citenamefont{{Y. Shao et al.}}(2014)}]{QChem}
\bibinfo{author}{\bibnamefont{{Y. Shao et al.}}}, \bibinfo{journal}{Mol. Phys.}
   (\bibinfo{year}{2014}).

\bibitem[{\citenamefont{{SymPy Development Team}}(2014)}]{SymPy}
\bibinfo{author}{\bibnamefont{{SymPy Development Team}}},
  \emph{\bibinfo{title}{SymPy: Python library for symbolic mathematics}}
  (\bibinfo{year}{2014}), \urlprefix\url{http://www.sympy.org}.

\bibitem[{\citenamefont{Jones et~al.}(2001)\citenamefont{Jones, Oliphant,
  Peterson et~al.}}]{SciPy}
\bibinfo{author}{\bibfnamefont{E.}~\bibnamefont{Jones}},
  \bibinfo{author}{\bibfnamefont{T.}~\bibnamefont{Oliphant}},
  \bibinfo{author}{\bibfnamefont{P.}~\bibnamefont{Peterson}},
  \bibnamefont{et~al.}, \emph{\bibinfo{title}{{SciPy}: Open source scientific
  tools for {Python}}} (\bibinfo{year}{2001}),
  \urlprefix\url{http://www.scipy.org/}.

\bibitem[{\citenamefont{Hunter}(2007)}]{Hunter:2007}
\bibinfo{author}{\bibfnamefont{J.~D.} \bibnamefont{Hunter}},
  \bibinfo{journal}{Computing In Science \& Engineering}
  \textbf{\bibinfo{volume}{9}}, \bibinfo{pages}{90} (\bibinfo{year}{2007}).

\bibitem[{\citenamefont{Head-Gordon et~al.}(1998)\citenamefont{Head-Gordon,
  Maslen, and White}}]{HeadGordonWhite_98JCP}
\bibinfo{author}{\bibfnamefont{M.}~\bibnamefont{Head-Gordon}},
  \bibinfo{author}{\bibfnamefont{P.~E.} \bibnamefont{Maslen}},
  \bibnamefont{and} \bibinfo{author}{\bibfnamefont{C.~A.} \bibnamefont{White}},
  \bibinfo{journal}{J. Chem. Phys.} \textbf{\bibinfo{volume}{108}},
  \bibinfo{pages}{616} (\bibinfo{year}{1998}).

\bibitem[{\citenamefont{Hehre et~al.}(1969)\citenamefont{Hehre, Stewart, and
  Pople}}]{HehrePople_69JCP}
\bibinfo{author}{\bibfnamefont{W.~J.} \bibnamefont{Hehre}},
  \bibinfo{author}{\bibfnamefont{R.~F.} \bibnamefont{Stewart}},
  \bibnamefont{and} \bibinfo{author}{\bibfnamefont{J.~A.} \bibnamefont{Pople}},
  \bibinfo{journal}{J. Chem. Phys.} \textbf{\bibinfo{volume}{51}},
  \bibinfo{pages}{2657} (\bibinfo{year}{1969}).

\end{thebibliography}
\end{document}